\documentclass{PoS}

\title{Radiative $B$ Decays at Belle}

\ShortTitle{Radiative $B$ Decays at Belle}

\author{\speaker{Akimasa Ishikawa}\\
        Tohoku University\\
        E-mail: \email{akimasa@epx.phys.tohoku.ac.jp}}

\abstract{We report recent results on radiative $B$ decays at Belle at the KEKB collider.}

\FullConference{The 39th International Conference on High Energy Physics (ICHEP 2018)\\
                4-11 July, 2018\\
		Seoul, Korea}

\def\Mbc{M_{\rm bc}}

\begin{document}

\section{Introduction}
$B$ meson decays via loop diagrams are sensitive probe to new physics~(NP). These processes in the SM are suppressed by the Cabibbo-Kobayashi-Maskawa matrix elements, $V_{ts}$ or $V_{td}$, and a loop factor. Since unobserved heavy particles from some NP models can contributes the decays, the branching fractions, direct $CP$ violation~($A_{CP}$), photon polarization, isospin asymmetry~($\Delta_{0+}$ or $\Delta_{0-}$) and difference of $A_{CP}$ between charged and neutral $B$ mesons~($\Delta A_{CP}$) might differ from the SM predictions. Radiative $B$ decays $b \to s \gamma$, which are mediated by loop diagrams, are experimentally and theoretically clean due to final states having a color singlet photon. Thus these are ideal tools to search for NP.

We report the measurements of radiative $B$ decays with a full data sample of 711~fb${}^{-1}$ accumulated by the Belle detector at the KEKB energy-asymmetric collider. 

\section{Evidence for Isospin Violation in $B \to K^* \gamma$~\cite{Horiguchi:2017ntw}}
We reconstruct $B^0 \to K^{*0} \gamma$ and $B^+ \to K^{*+} \gamma$ decays, where $K^*$ is formed from $K^+ \pi^-$, $K_S^0 \pi^0$, $K^+ \pi^0$ or $K_S^0 \pi^+$ combinations. 
To determine the branching fractions and direct $CP$ asymmetries as well as $\Delta A_{CP}$ and isospin asymmetry~($\Delta_{0+}$), we perform extended unbinned maximum likelihood fits to the seven $\Mbc$ distributions and the results are
  \begin{eqnarray*}
{\cal{B}}(B^0 \to K^{*0} \gamma) &=& (3.96 \pm 0.07 \pm 0.14)\times10^{-5},\\ 
{\cal{B}}(B^+ \to K^{*+} \gamma) &=& (3.76 \pm 0.10 \pm 0.12)\times10^{-5},\\ 
A_{CP}(B^0 \to K^{*0} \gamma) &=& (-1.3 \pm 1.7 \pm 0.4)\%,\\ 
A_{CP}(B^+ \to K^{*+} \gamma) &=& (+1.1 \pm 2.3 \pm 0.3)\%,\\ 
A_{CP}(B \to K^{*} \gamma)    &=& (-0.4 \pm 1.4 \pm 0.3)\%,\\ 
\Delta_{0+}   &=& (+6.2 \pm 1.5 \pm 0.6 \pm 1.2)\%, \\
\Delta A_{CP} &=& (+2.4 \pm 2.8 \pm 0.5)\%, 
  \end{eqnarray*}
where the third uncertainty for $\Delta_{0+}$ is due to the uncertainty in $f_{+-}/f_{00}$. 
We find evidence for isospin violation in $B \to K^* \gamma$ decays with a significance of 3.1$\sigma$, and this result is consistent with the predictions in the SM. Measured branching fractions and $A_{CP}$ are most precise to date.



\section{Measurement of Time-Dependent $CP$ Violation in $B \to K \eta \gamma$~\cite{Nakano:2018lqo}}
A measurement of time-dependent $CP$ violation in ${B^0 \to P_1^0 P_2^0 \gamma}$ is the most promising method to measure the photon polarization in the $b \to s \gamma$ process, where $P_1^0$ and $P_2^0$ are scalar or pseudoscalar mesons and the $P_1^0P_2^0$ system is a $CP$ eigenstate. We  measure the time-dependent $CP$ violation in ${B^0 \to K_S^0 \eta \gamma}$.
We determine $CP$ violation parameters, ${\cal S}$ and ${\cal A}$, by performing an unbinned maximum-likelihood fit to the observed proper time difference distribution in the signal region. The obtained parameters
\begin{eqnarray*}
\nonumber
    {\cal S} &=& -1.32 \pm 0.77 {\rm (stat.)} \pm 0.36{\rm (syst.)}, \\
\nonumber
    {\cal A} &=& -0.48 \pm 0.41 {\rm (stat.)} \pm 0.07{\rm (syst.)}
\end{eqnarray*}
are consistent with the null-asymmetry hypothesis within 2$\sigma$ as well as with SM predictions.

\section{Measurements of Isospin Asymmetry and Difference of $CP$ Asymmetries in $B \to X_s \gamma$~\cite{Watanuki:2018xxg}}
Precision measurements of $B \to X_s \gamma$ branching fraction~${\cal{B}}(B \to X_s \gamma)$
~\cite{Limosani:2009qg,Saito:2014das}
and direct $CP$ asymmetry~\cite{Nishida:2003paa} 
are in good agreement with the SM predictions~\cite{Misiak:2015xwa,Benzke:2010tq} and set strong constraints on NP models. The theoretical uncertainties in the predictions of ${\cal{B}}(B \to X_s \gamma)$ and $A_{CP}(B \to X_s \gamma)$ are about 7\% and 2\% which are comparable with the experimental uncertainties of the current world averages~\cite{Tanabashi:2018oca}. 
The Belle~II experiment is expected to measure the branching fraction with a precision of about 3\%~\cite{Kou:2018nap}. Thus, the reduction of the theoretical uncertainty is crucial to further constrain NP models. One of the largest theoretical uncertainties is a resolved photon contribution which can be constrained from the measurement of isospin asymmetry in $B \to X_s \gamma$~\cite{Lee:2006wn,Misiak:2009nr,Benzke:2010js}.
A newly proposed $CP$ violation observable is a difference of the direct $CP$ asymmetries between the charged and neutral $B$ mesons, which is proportional to a ratio of Wilson coefficients ${\rm Im}(C_8/C_7)$, is null in the SM while in some NP models, this value could be as large as 10\%~\cite{Benzke:2010tq,Malm:2015oda,Endo:2017ums}.

We perform measurements of $\Delta_{0-}$ and $\Delta A_{CP}$ in $B \to X_s \gamma$ with sum of 38 exclusive $X_s$ final states which cover about 77\% of total $X_s$ decay rate.
To extract the physics observables, we perform a simultaneous fit with an extended unbinned maximum likelihood method to eight $M_{\rm bc}$ distributions; five for $B^-$, $B^+$, $\bar{B}^0$, $B^0$, and $B_{\rm fns}$ in the on-resonance data, and three for charged $B$~($B^-$ and $B^+$), flavor-specific neutral $B$~($\bar{B}^0$ and $B^0$), and $B_{\rm fns}$ in the off-resonance data, where $B_{\rm fns}$ is denoted as flavor non-specific final states. The results are 
\begin{eqnarray}
\nonumber
  \Delta_{0-}   &=& (-0.48 \pm 1.49  \pm 0.97  \pm 1.15)\%,\\
\nonumber
  \Delta A_{CP} &=& (+3.69 \pm 2.65  \pm 0.76)\%,\\
\nonumber
  A_{CP}^{\rm C}    &=& (+2.75 \pm 1.84  \pm 0.32)\%, \\
\nonumber
  A_{CP}^{\rm N}    &=& (-0.94 \pm 1.74  \pm 0.47)\%,\\
\nonumber
  A_{CP}^{\rm tot}  &=& (+1.44 \pm 1.28  \pm 0.11)\%,\\
\nonumber
  \bar{A}_{CP} &=& (+0.91 \pm 1.21  \pm 0.13)\%,
\end{eqnarray}
where the third uncertainty for $\Delta_{0-}$ is due to $f_{+-}/f_{00}$.
The measured $\Delta_{0-}$ and $\Delta A_{CP}$ are consistent with zero. Thus, these measurements can be used to constrain the resolved photon contribution in $B \to X_s \gamma$ and to constrain NP models. All the measurements are most precise to date.

\section{Summary}
We have measured the branching fractions, isospin asymmetries and $CP$ violation
observables in radiative $B$ decays at Belle. These measurements strongly constrain NP models.
Our measurements of $CP$ violation observables are dominated by
the statistical uncertainty; thus, the upcoming Belle~II
experiment will further reduce the uncertainty. To improve the isospin asymmetry
at Belle~II, reduction of the dominant uncertainty due to $f_{+-}/f_{00}$ is essential, and can be performed at both Belle and Belle~II.

\section*{Acknowledgments}
A.~I. is supported by the Japan Society for the Promotion of Science (JSPS) Grant No.~16H03968.

\end{document}